\documentclass{appolb}
\usepackage{epsfig}

\begin{document}
\title{Pattern formation and spatial correlation induced by the noise in two competing species%
\thanks{Presented at the $16^{th}$ Marian Smoluchowski Symposium on Statistical Physics:
Fundamentals and Applications, Zakopane, Poland, September 6-11,
2003
}%
}
\author{Valenti D., Fiasconaro A., Spagnolo B.
\address{INFM - Unit\`a di Palermo, Group of Interdisciplinary Physics and\\
Dipartimento di Fisica e Tecnologie Relative, Universit\`a di
Palermo\\
Viale delle Scienze, I-90128 Palermo, Italy} } \maketitle
\begin{abstract}
We analyze the spatio-temporal patterns of two competing species
in the presence of two white noise sources: an additive noise
acting on the interaction parameter and a multiplicative noise
which affects directly the dynamics of the species densities. We
use a coupled map lattice (CML) with uniform initial conditions.
We find a nonmonotonic behavior both of the pattern formation and
the density correlation as a function of the multiplicative noise
intensity.
\end{abstract}



\section{Introduction}
We present a stochastic model for spatial distribution of two
competing species. Our theoretical model could be useful to
describe biological systems, where the presence of fluctuations,
such as random variability of temperature, can modify strongly the
dynamics of an ecosystem~\cite{Zimmer,Ciuchi}. We focus on the
role played by the noise on the transient dynamics of the spatial
distributions of two competing species belonging to an ecosystem
described by generalized Lotka-Volterra equations~\cite{Lotka} in
the presence of multiplicative noise. We find nonmonotonic
behaviors for the pattern formation and the density correlation of
the species as a function of the multiplicative noise intensity.
The theoretical results could contribute to select environmental
and periodical driving forces using the proper space and time
scales to develop physical models of population dynamics useful to
interpret spatial patterns in the abundance of the
species~\cite{Mazzola,Garcia}.

\section{The model}
To study the spatial effects due to the presence of noise sources
we consider a discrete time evolution model, which is the discrete
version of the Lotka-Volterra equations with diffusive terms,
namely a coupled map lattice~\cite{Kaneko}.
\begin{figure}[htbp]
\begin{center}
\includegraphics[width=9cm]{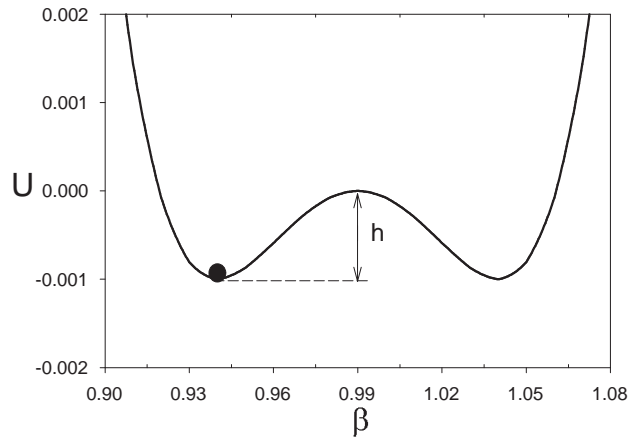}
\end{center}
\caption{ \small \emph{The bistable potential $U(\beta)$ of the
interaction parameter $\beta(t)$. The potential $U(\beta)$ is
centered on $\beta=0.99$. The parameters of the potential are
 $h = 6.25 \cdot 10^{-3}$, $\eta=0.05$, $\rho = -0.01$.}\bigskip}
\label{potential}
\end{figure}
The time evolution of the spatial distribution for the two species
is given by the following equations
\begin{eqnarray}
x_{i,j}^{n+1}&=&\mu x_{i,j}^n (1-x_{i,j}^n-\beta^n
y_{i,j}^n)+\sqrt{\sigma_x}
x_{i,j}^n X_{i,j}^n + D\sum_\gamma (x_{\gamma}^n-x_{i,j}^n),
\label{Lotka_eq_1}\\
y_{i,j}^{n+1}&=&\mu y_{i,j}^n (1-y_{i,j}^n-\beta^n
x_{i,j}^n)+\sqrt{\sigma_y} y_{i,j}^n Y_{i,j}^n + D\sum_\gamma
(y_{\gamma}^n-y_{i,j}^n), \label{two} \label{Lotka_eq_2}
\end{eqnarray}
where $x^n_{i,j}$ and $y^n_{i,j}$ denote respectively the
densities of prey 1 and prey 2 in the site $(i,j)$ at the time
step $n$, $\mu$ is proportional to the growth rate, $D$ is the
diffusion constant, $\sum_\gamma$ indicates the sum over the four
nearest neighbors.
\begin{figure}[htbp]
\begin{center}
\includegraphics[width=9cm]{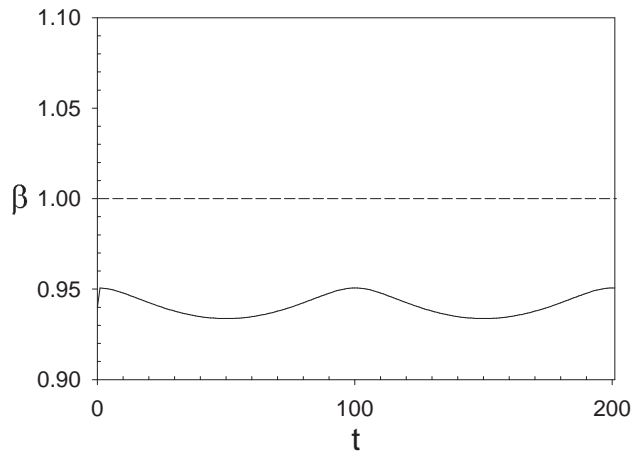}
\end{center}
\caption{\small \emph{Time evolution of the interaction parameter
$\beta$ for $\sigma_\beta=0$ and initial value $\beta(0)=0.94$.
The values of the parameters are: $\gamma=1.5 \cdot 10^{-1}$,
$\omega_0/(2\pi)=10^{-2}$. In the absence of noise $\beta(t)$
oscillates below the critical value $\beta_c=1$. The transient
behavior near the initial point $\beta(0)=0.94$ is due to the
choice of cosine function in eq.(\ref{beta_eq}) as deterministic
periodical driving.}\bigskip} \label{beta_1}
\end{figure}
The random terms are modeled by independent Gaussian variables
denoted by $X^n_{i,j}$, $Y^n_{i,j}$ with zero mean and variance
unit and $\beta^n$ takes into account for the interaction between
the species. In eqs.~(\ref{Lotka_eq_1}) and (\ref{Lotka_eq_2}),
$\sigma_x$ and $\sigma_y$ are the intensities of the
multiplicative noise which model the interaction between the
species and the environment.

\subsection{Stochastic Resonance}

It is known that for $\beta < 1$ a coexistence regime takes place,
that is both species survives, while for $\beta > 1$ an exclusion
regime is established, that is one of the two species vanishes
after a certain time~\cite{Vilar,Spagnolo}. Coexistence and
exclusion of one of the two species correspond to stable states of
the Lotka-Volterra's deterministic model~\cite{Lotka}. Real
ecosystems are immersed in a noisy nonstationary environment, so
also the interaction parameter is affected by the noise and some
other deterministic periodical driving such as the temperature.
The change in the competition rate between exclusion and
coexistence occurs randomly because of the coupling between the
limiting resources and the noisy environment. A random variation
of limiting resources produces a random competition between the
species. The noise therefore together with the periodic force
determines the crossing from a dynamical regime ($\beta < 1$,
coexistence) to the other one ($\beta > 1$, exclusion). To
describe this continuous and noisy behaviour of the interaction
parameter $\beta(t)$ we consider an Ito stochastic differential
equation with a bistable potential and a periodical driving force
\begin{equation}
\frac{d\beta(t)}{dt} = -\frac{dU(\beta)}{d\beta}+\gamma
cos(\omega_0 t) + \xi_{\beta}(t) \label{beta_eq},
\end{equation}
where $U(\beta)$ is the bistable potential (see
Fig.(\ref{potential}))
\begin{equation}
U(\beta) = h(\beta-(1+\rho))^4/\eta^4-2h(\beta-(1+\rho))^2/\eta^2,
\label{U(beta)}
\end{equation}
the periodical driving mimics the climatic temperature
oscillations, and $\xi_\beta(t)$ is a white Gaussian noise with
$<\xi_\beta(t)>=0$ and
$<\xi_\beta(t)\xi_\beta(t')>=\sigma_\beta\thinspace \delta(t-t')$.
Since the dynamics of the species strongly depends on the value of
the interaction parameter, we initially consider the time
evolution of $\beta$ for $\sigma_\beta=0$ (see
Fig.(\ref{beta_1})).
\begin{figure}[htbp]
\begin{center}
\includegraphics[width=9cm]{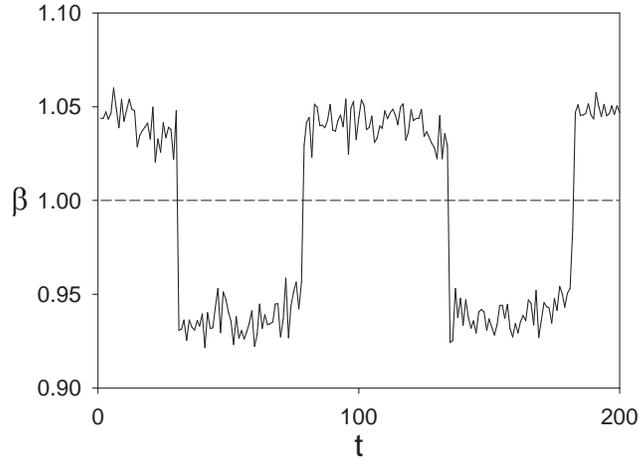}
\end{center}
\caption{ \small \emph{Time evolution of the interaction parameter
$\beta(t)$ for $\sigma_\beta = 2.65 \cdot 10^{-3}$. The values of
the parameters and the initial value $\beta(0)$ are the same of
Fig.\ref{beta_1}. For this level of noise a synchronization
appears (Stochastic Resonance) and $\beta(t)$ oscillates
quasi-periodically below and above the critical value $\beta_c=1$.
}\bigskip} \label{beta_2}
\end{figure}
We note that in the absence of the additive noise $\xi_\beta(t)$,
$\beta(t)$ shows a periodical evolution but its values always
remain below $\beta = 1$, i. e. in the coexistence regime. The
noise can synchronize with the periodical driving force. In this
case a Stochastic Resonance (SR)~\cite{Benzi, Alley} effect
appears which affects strongly the dynamics of the
system~\cite{Vilar,Spagnolo1,Valenti1}. Therefore we fix the
additive noise intensity at the value $\sigma_\beta=2.65 \cdot
10^{-3}$ corresponding to a competition regime with $\beta$
periodically switching from coexistence to exclusion
regions~\cite{Valenti1} (see Fig.\ref{beta_2}). The SR effect in
the dynamics of interaction parameter $\beta$ induces SR
phenomenon in the dynamics of two competing species. This produces
noise-induced anticorrelated periodic oscillations in the time
evolution of the two species densities (see ref.~\cite{Valenti1}.

\subsection{Spatial distributions}

\begin{figure}[htbp]
\begin{center}
\includegraphics[width=12cm]{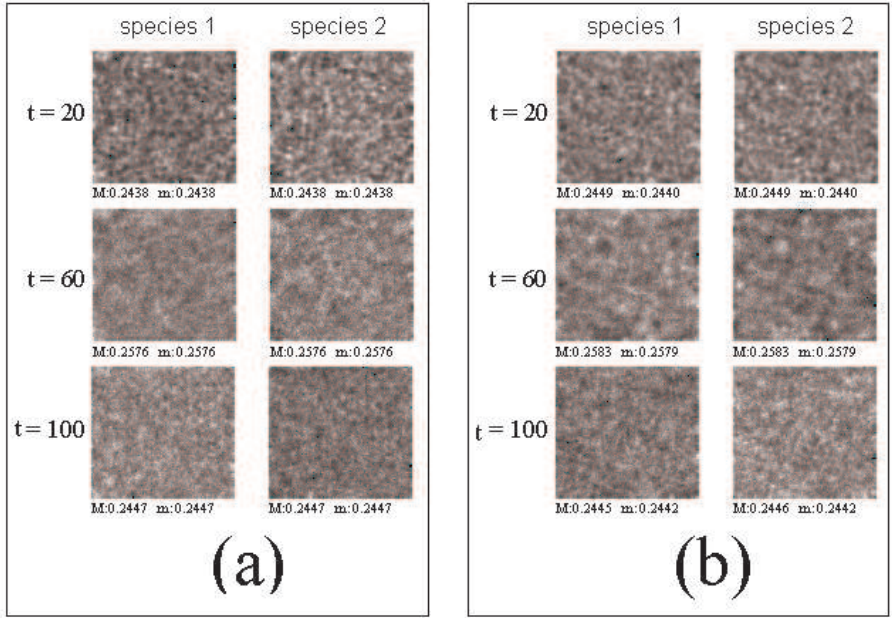}
\end{center}
\caption{\small\emph{Spatial distributions at different times for
(a) $\sigma= 10^{-12}$ and (b) $\sigma= 10^{-8}$. The value of the
additive noise is fixed at $\sigma_\beta = 2.65 \cdot 10^{-3}$.
The values of the parameters are: $\mu=2$, $D=0.05$, $\gamma=1.5
\cdot 10^{-1}$, $\omega_0/(2\pi)=10^{-2}$, $N = 100\times 100$.
The initial values are $x^0_{i,j}=y^0_{i,j}=0.5$ for all sites
$(i,j)$ and $\beta(0)= 0.94$.}\bigskip} \label{spatial1}
\end{figure}

\begin{figure}[htbp]
\begin{center}
\includegraphics[width=12cm]{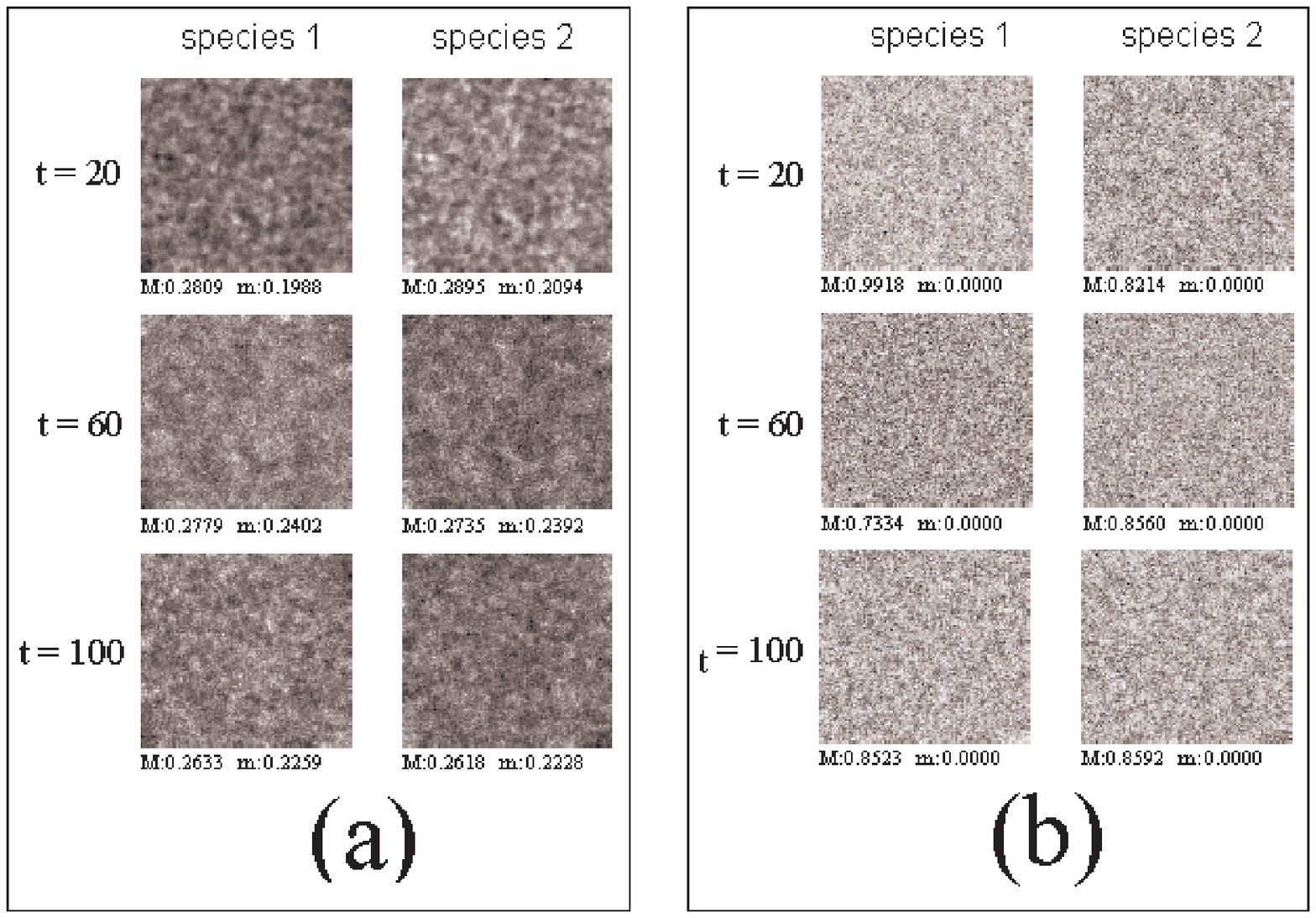}
\end{center}
\caption{ \small \emph{Spatial distributions at different times
for (a) $\sigma= 10^{-4}$ and (b) $\sigma= 10^{-1}$. The additive
noise intensity, the values of the parameters and the initial
conditions are the same of Fig.\ref{spatial1}.}\bigskip}
\label{spatial2}
\end{figure}

We consider the time evolution of the spatial distribution of the
ecosystem, described by Eqs. (\ref{Lotka_eq_1}) and
(\ref{Lotka_eq_2}), in the SR dynamical regime. The interaction
parameter $\beta^n$ of eqs.(\ref{Lotka_eq_1}), (\ref{Lotka_eq_2})
corresponds to the value of continuous $\beta(t)$ of
eq.(\ref{beta_eq}) taken at the step $n$. So we fix the additive
noise at the value $\sigma_\beta=2.65 \cdot 10^{-3}$ and we vary
both the intensities of multiplicative noise. In
Figs.~\ref{spatial1} and \ref{spatial2}
 we report the spatio-temporal patterns of the two
species for different values of the multiplicative noise intensity
$\sigma = \sigma_x = \sigma_y$, namely $\sigma = 10^{-12},
10^{-8}, 10^{-4}, 10^{-1}$ with $\mu = 2$, $D = 0.05$, $A = 1.5
\cdot 10^{-1}$, $\omega_0/(2\pi) = 10^{-2}$, $\beta(0)=0.94$ and
$x^0_{i,j}=y^0_{i,j}=0.5$ at all sites $(i,j)$. We see that for
very low noise intensity (see Fig.\ref{spatial1}a) an average
correlation on the considered lattice ($N = 100 \times 100$)
between the species is observed. For higher noise intensities (see
Fig.\ref{spatial1}b, \ref{spatial2}a) an anticorrelation between
the two species is observed: the two species tend to occupy
different positions. The anticorrelation is more evident in
Fig.\ref{spatial2}a. By increasing the multiplicative noise the
anticorrelation is strongly reduced (see Fig.\ref{spatial2}b).
Further increase of the noise ($\sigma = 10^{+3}$) causes the
anticorrelation to disappear and the two species densities become
uncorrelated. We note also that the average size of the patterns
increases with the noise intensity non-monotonically: at very low
noise intensity ($\sigma = 10^{-14}$) the spatial distribution is
almost uniform, by increasing the noise intensity spatial patterns
arise (see Figs.\ref{spatial1}b, \ref{spatial2}a) and a further
increase of the noise intensity reduces the average size of the
patterns (see Fig.\ref{spatial2}b).

\subsection{Spatial correlation}

\begin{figure}[htbp]
\begin{center}
\includegraphics[width=12cm]{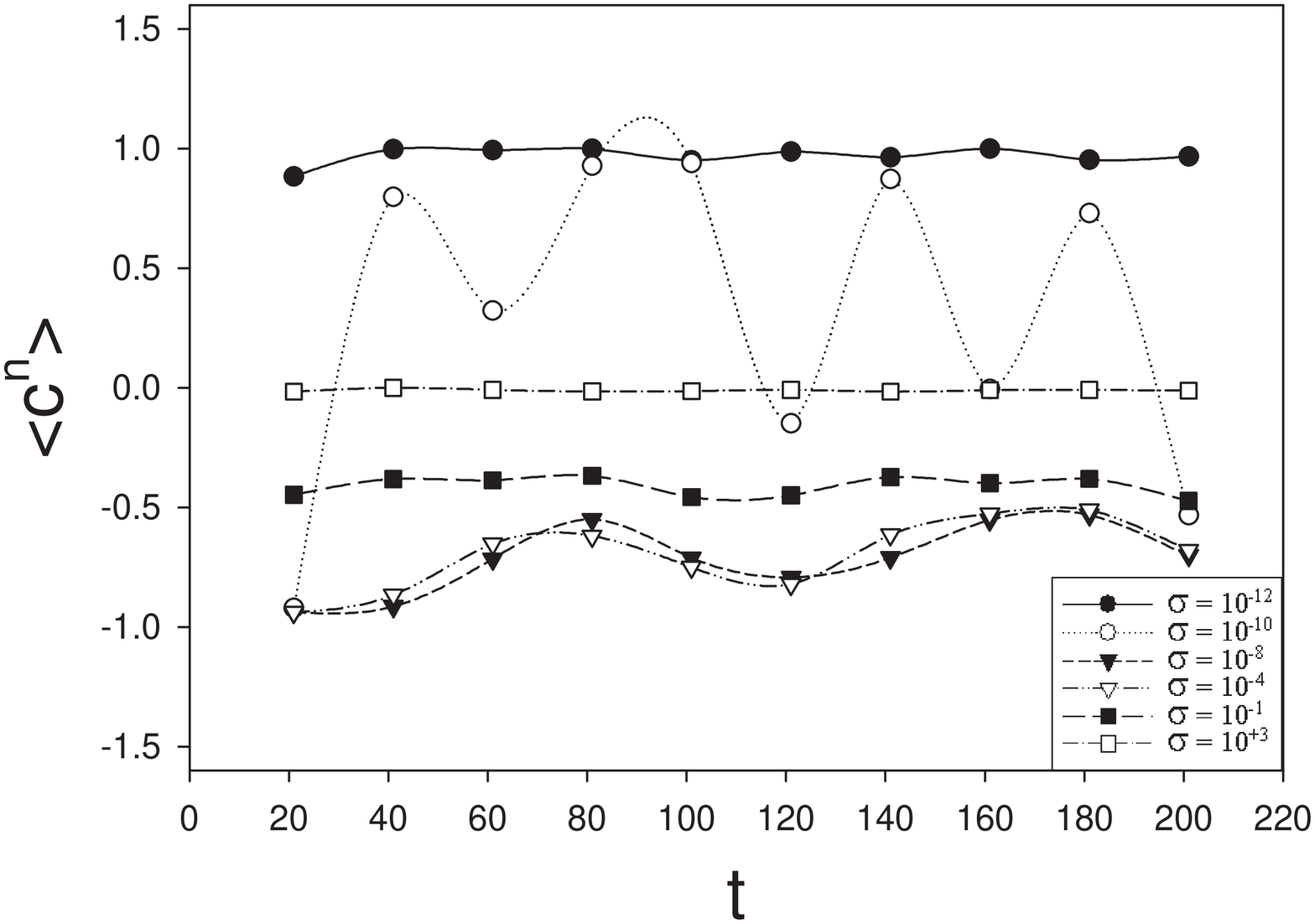}
\end{center}
\caption{ \small \emph{Mean correlation coefficient $<c^n>$ as a
function of the time. For low levels of the multiplicative noise
($\sigma = 10^{-12}$) the species are strongly correlated and
$<c^n>$ is approximately constant. By increasing the intensity of
the multiplicative noise ($\sigma = 10^{-10}$) $<c^n>$ shows big
fluctuations. A further increase of the noise ($\sigma = 10^{-8}$,
$\sigma = 10^{-4}$) causes strong anticorrelation between the two
species with $<c^n>$ oscillating at the frequency of the
periodical forcing. For very high levels of noise, the
anticorrelation is reduced ($\sigma = 10^{-1}$) and finally it
disappears ($\sigma = 10^{+3}$), that is the species are totally
uncorrelated.}\bigskip}
 \label{MCC}
\end{figure}

In order to evaluate the spatial correlation between the two
species for the noise intensities considered we calculate, at the
time step $n$, the mean correlation coefficient $<c^n>$ defined on
the lattice as

\begin{equation}
<c^n>=\frac{cov^n_{xy}}{s^n_x s^n_y} \label{correlation}
\end{equation}
with

\begin{equation}
cov^n_{xy}=\frac{\sum_{i,j}(x^n_{i,j}-\bar{x}^n)(y^n_{i,j}-\bar{y}^n)}{N}
\label{covariance},
\end{equation}
where $\bar{x}^n$, $s^n_x$, $\bar{y}^n$, $s^n_y$ are the mean
value and the root mean square respectively of species 1 and
species 2, obtained over the whole spatial grid at the time step
$n$, $cov^n_{xy}$ is the corresponding covariance and $N =
100\times 100$ the number of sites which compose the grid. The
behaviour of the mean correlation coefficient $<c^n>$ as a
function of the time for different levels of the multiplicative
noise has been reported in Fig.\ref{MCC}.
\begin{figure}[htbp]
\begin{center}
\includegraphics[width=10cm]{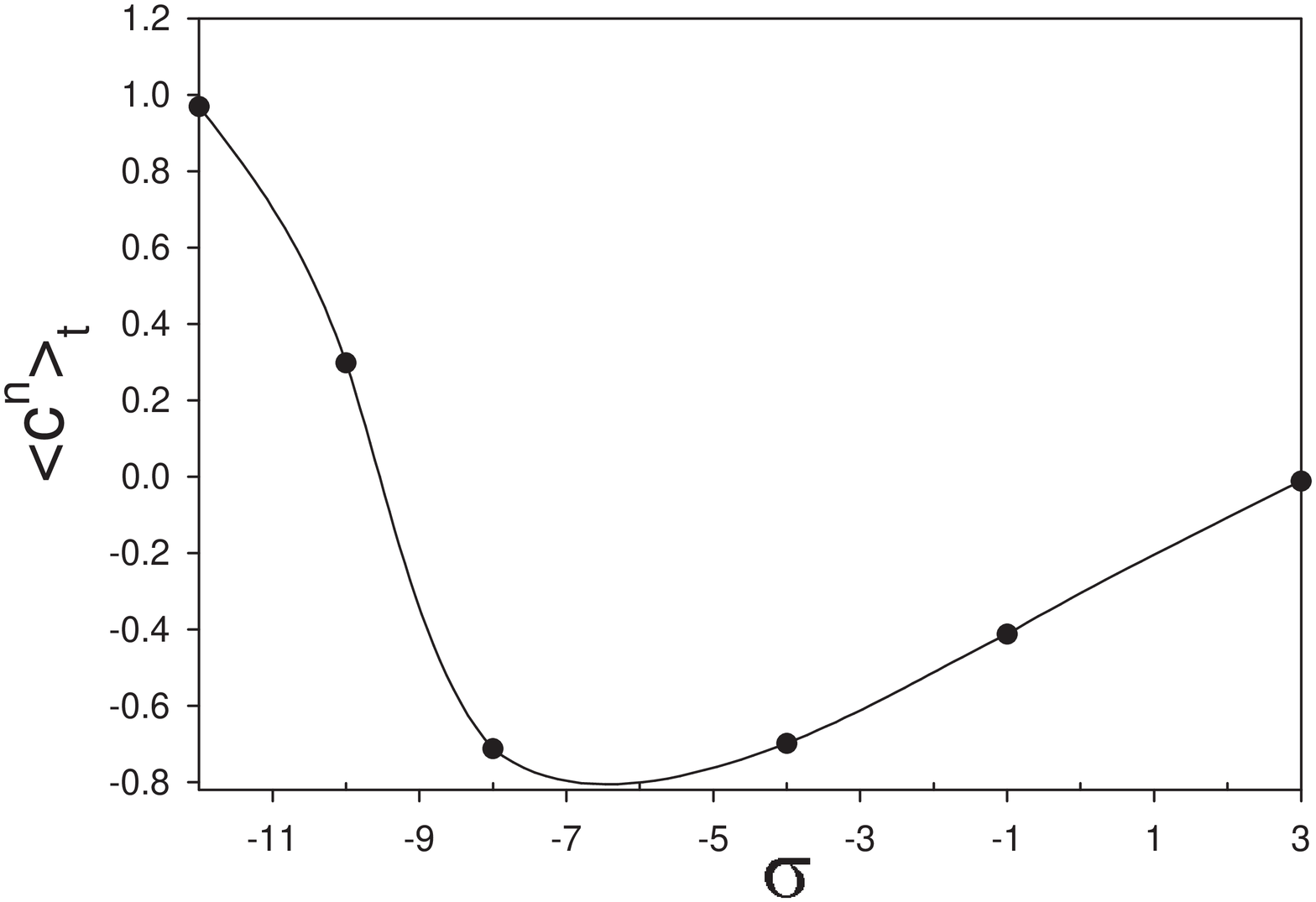}
\end{center}
\caption{ \small \emph{Time average of the mean correlation
coefficient $<c^n>_t$ as a function of the multiplicative noise in
semilog scale.}\bigskip} \label{MCC_aver}
\end{figure}
We observe a nonmonotonic behaviour of $<c^n>$ as a function of
the multiplicative noise intensity. In fact for low noise
intensities $\sigma = 10^{-12}$, $<c^n>$ shows weak oscillation
around $1$, that is strong correlation between the two species.
For higher levels of the noise $\sigma = 10^{-10}$, $<c^n>$ is
affected by fluctuations and its values vary strongly as a
function of the time. A further increase of the multiplicative
noise, i.e. $\sigma = 10^{-8}$ and $\sigma = 10^{-4}$, determines
an oscillation of $<c^n>$ around a negative value, that is
anticorrelation between the two species, with the frequency of the
periodical forcing. For higher intensities of the noise $\sigma =
10^{-1}$, the value of the mean correlation coefficient $<c^n>$
increases and it vanishes for $\sigma = 10^{+3}$. Finally to show
clearly the nonmonotonic behaviour of $<c^n>$, we calculate the
time average of the mean correlation coefficient $<c^n>_t$ and we
report it as a function of the multiplicative noise intensity in
Fig.\ref{MCC_aver}. A clear minimum is shown, which corresponds to
the anticorrelated oscillations shown in the time evolution of two
competing species in each point of our spatial grid
\cite{Valenti1}. We note therefore the different role of the two
noise sources in the ecosystem dynamics. The additive noise
determines the conditions of the dynamical regime, the
multiplicative noise produces a coherent response of the system
(see ref.\cite{Valenti1}), which is responsible for the appearence
of anticorrelation behavior in the spatial patterns of the
species.

\section{Conclusions}

We present a study on the role of the noise in the spatial
distributions of two interacting species. The main result is that
noise plays a very important role in population dynamics and
cannot be neglected. Noise can have a constructive role and it is
responsible for the enhancement of the response of the system to a
driving force producing stochastic resonance. By using a discrete
time evolution model, which is the discrete version of the
Lotka-Volterra equations with diffusive terms, in the presence of
a multiplicative noise and with a random interaction parameter, we
analyze the temporal behaviours of the spatial distributions for
an ecosystem consisting of two species. The noise induces
spatio-temporal behaviors which are absent in the deterministic
dynamics, i.e. pattern formation with the same periodicity of the
deterministic force. Moreover appearance of temporal oscillation
is observed in the correlation coefficient between the two
species. We find a nonmonotonic behaviour of the time average
correlation coefficient as a function of the multiplicative noise.
In fact at low levels of the noise intensity the species densities
are almost uniform and a strong correlation appears in the spatial
distributions. By increasing the noise intensity we observe
pattern formation and anticorrelated behaviour, which exhibit the
same periodicity of the deterministic driving force and
corresponds to the minimum of Fig.\ref{MCC_aver}. For higher
values of the noise intensity no patterns or correlations appear.
Our model could be useful to explain spatio-temporal behaviours of
populations, whose dynamics is strongly affected by the noise and
by the environmental physical variables.

\end{document}